\documentclass[12pt,amssymb,epsfig]{article}
\usepackage{latexsym}
\usepackage{amsfonts}
\input  epsf 
\newtheorem{theo}{Theorem}

\newenvironment{proof}{{\bf Proof} }{$$ \eqno \Box$$}

\def\rlx{\relax\leavevmode}
\def\inbar{\vrule height1.5ex width.4pt depth0pt}
\def\IZ{\rlx\hbox{\small \sf Z\kern-.4em Z}}
\def\IR{\rlx\hbox{\rm I\kern-.18em R}}
\def\ID{\rlx\hbox{\rm I\kern-.18em D}}
\def\IC{\rlx\hbox{\,$\inbar\kern-.3em{\rm C}$}}
\def\zero{\rlx\hbox{\,$\inbar\kern-.18em{ 0}$}}
\def\IN{\rlx\hbox{\rm I\kern-.18em N}}
\def\one{\hbox{{1}\kern-.25em\hbox{l}}}
\def\slash#1{#1 \!\!\! /}

\def\beq{\begin{equation}}
\def\eeq{\end{equation}}
\def\bea{\begin{eqnarray}}
\def\eea{\end{eqnarray}}
\def\ber{\begin{array}}
\def\eer{\end{array}}

\begin{document}

\begin{titlepage}

March 2001 \hfill{UTAS-PHYS-01-02}\\
\vskip 1.6in
\begin{center}
{\Large {\bf Algebraic solution for the vector potential in the
Dirac equation}}\end{center}

\normalsize
\vskip .4in

\begin{center}
H S Booth$^{*}$, \hspace{3pt} G Legg, \hspace{3pt}
P D Jarvis
\par \vskip .1in \noindent
{\it School of Mathematics and Physics, University of Tasmania}\\
{\it GPO Box 252-21, Hobart Tas 7001, Australia }\\

\end{center}
\par \vskip .3in

\begin{center}
{\large {\bf Abstract}}\\
\end{center}
The Dirac equation for an electron in an external electromagnetic field can be regarded as a 
singular set of linear equations for the vector potential.
Radford's method of algebraically solving for the vector potential is reviewed, with 
attention to the additional constraints arising from non-maximality of 
the rank. The extension of the method to general spacetimes is 
illustrated by  examples in diverse dimensions with both 
$c$- and 
$a$-number wavefunctions. 

\vfill
\noindent
\footnotesize $^{*}$Centre for Mathematics and its Applications, Australian National 
University.
\texttt{hbooth@wintermute.anu.edu.au}
\end{titlepage}

\section{Introduction}
The Maxwell-Dirac equations are the coupled nonlinear partial 
differential equations which describe a classical electron interacting 
with an electromagnetic field. They are also the equations from which 
quantum electrodynamics is derived. Since the mathematical 
foundations of the latter remain unclear, the Maxwell-Dirac equations 
continue to be of interest \cite{Esteban,Flato,Georgiev}. Recently 
Radford\cite{Radford96} handled the Maxwell-Dirac 
equations firstly by solving the Dirac equation for the 
electromagnetic potential in terms of the wavefunction and its 
derivatives, and then substituting this solution in the Maxwell equations. 
This approach subsequently led to some physically interesting 
results\cite{BoothRadford,RadfordBooth,Booth98} 
(for a review see \cite{Booth00}). 

Despite the viability and potential importance of Radford's algebraic solution,
at least for the treatment of the equations of classical 
electrodynamics for electrons and photons, it appears that 
the method has not before appeared in this context. 
One analysis which reached negative conclusions about the approach, and which may have 
engendered the lack of attention to it in the literature, is that of 
Eliezer\cite{Eliezer}. In that paper, it was noted that the determinant of 
the matrix of coefficients of the vector potential $A_{\mu}$ in the 
Dirac equation actually vanishes, and that therefore a umique algebraic 
inversion was \textit{not} possible.
The aim of the present paper is to reconcile \cite{Radford96} and \cite{Eliezer}, and 
to emphasise the legitimacy of the algebraic \textit{Ansatz}, despite the 
negative conclusions of \cite{Eliezer}, by a careful analysis of the nature of the 
Dirac equation regarded as a linear system\cite{LinearAlg} for $A_{\mu}$. The main 
result is that the Dirac equation is indeed invertible if a \textit{real}
solution for the vector potential is required, and moreover that the 
treatment entrains an additional set of polynomial constraints on the wavefunction and its
partial derivatives which must be carried forward in any further 
analysis. In \S 2 below, the abstract formalism is developed, and (for the 4 dimensional case)
it is shown how the explicit manipulations, which rely on the structure of 
the Dirac algebra to derive the solution for the vector potential 
and the additional constraints, conform to the general setting (it is 
also pointed out that the solution can be regarded as including the 
mass, or more generally a Lorentz scalar potential, as a fifth unknown). This 
is done both in Lorentz-covariant Dirac spinor notation, and in van der 
Waerden 2-spinor notation. In \S 3, the case of arbitrary (flat) 
spacetimes with signature $(t,s)$ is taken up. Known results on the 
structure of the Dirac algebra (formally, the Clifford 
algebra ${\mathcal C}(t,s)$) in these cases are used to give an 
enumeration of constraints which are 
\textit{quadratic} in the wavefunction and derivatives
(in addition to current and partial axial 
current conservation, which hold in all cases). The 
four-dimensional results are recovered, and generalised to the 
case of $a$-number as well as $c$-number wavefunctions. A major 
outcome is a tabulation (table \ref{tbl:DiverseDimns}) of such
constraints as to fermion wavefunction statistics and metric 
signature in diverse dimensions. Concluding remarks and prospects for 
further development of the work are given in \S 4 below. 

\section{The 4 dimensional Dirac equation}
\subsection*{$8 \times 4$ real system}

The Dirac equation 
for a fermion of charge $q$ described by the spinor wavefunction $\psi$  
in the presence of an external electromagnetic potential may be written\footnote{
In this section standard Cartesian coordinates $x^{\mu}$, $\mu = 
0,1,2,3$ for four dimensional Minkowski space with $(1,3)$ metric $(\eta_{\mu 
\nu})_{\mu,\nu = 0,1,2,3} = \mbox{diag}(+,-,-,-)$ are introduced. 
Affices for Dirac spinors are introduced as $\psi_{\alpha}$, 
$\alpha = 1,2,3,4$, while the Dirac matrices (generators of the 
Clifford algebra ${\mathcal C}(1,3)$ in the standard basis) satisfy
$\{ \gamma_{\mu}, \gamma_{\nu} \} = 2 \eta_{\mu \nu}$; for 
conventions see \cite{ItzyksonZuber}. In \S 3 below,
the notation is generalised to dimension $d=t+s$ with metric 
signature $(t,s)$.} 
\begin{equation}
q \gamma^\mu \psi  A_\mu = (i\gamma ^\mu \partial_\mu \psi -m \psi ).
\label{eq:DiracEqn} 
\end{equation}
Following Eliezer\cite{Eliezer}, we write this as a matrix equation for $A_\mu$;
\begin{equation}
M_\alpha^\mu A_\mu =Z_\alpha
\label{eq:DiracMatrixForm} 
\end{equation}
where $M_\alpha^\mu \equiv {{\gamma^{\mu}}_\alpha}^{\beta} \psi _\beta$
for $M \in M_4({\mathbb C})$, $M:{\mathbb C}^4\longrightarrow {\mathbb C}^4$ and 
$A,\; Z \in {\mathbb C}^4$. 

In \cite{Eliezer} it was noted that $M$
had rank 3 and determinant
zero, with a rank 1 right null space,
and therefore could not be inverted to obtain a unique 
solution for the potential. Yet Radford\cite{Radford96} did just this, albeit in the bispinor 
representation. That work exploited the fact that $A_\mu$ is real\footnote{
The most telling use of the reality of $A_\mu$ 
is implicit in the bispinor representation, 
where half the equations are conjugated,
taking $A_\mu$ to be real.  This results in systems of equations where the 
matrix operating on $A_\mu$ can have non-zero determinant \cite{Booth98}.
This transformation cannot be performed with a matrix transformation 
on ${\mathbb C}^4$, and does not preserve the determinant.
}, which was not used in \cite{Eliezer}.
The point is that despite the zero determinant, (\ref{eq:DiracMatrixForm})
can be inverted if we know that $A_\mu$ is real, and providing that the 
intersection of the right null space of $M$ with ${\mathbb R}^4$ (as a 
subspace of ${\mathbb C}^4$ ) is trivial.  Even though the columns of
$M$ are not linearly independent as a vector space over ${\mathbb C}$,
they are in general linearly independent as a vector space over ${\mathbb R}$.

We may break (\ref{eq:DiracMatrixForm}) into real and imaginary parts,
yielding a system of 8 real equations in 4 real unknowns,
schematically ${\mathcal M} A = {\mathcal Z}$, where
\begin{equation} 
{\mathcal M} = \left( \begin{array}{c} \frac 12 (M+M^{*}) \\ \frac 1{2i}(M-M^{*})
\end{array} \right) , \quad  
{\mathcal Z} = \left( \begin{array}{c} \frac 12 (Z+Z^{*}) \\ \frac 
1{2i}(Z-Z^{*}) 
\end{array} \right).
\label{eq:8times4real}
\end{equation}
${\mathcal M}$ is not square; no determinant is defined; 
yet there are other tests for linear independence and 
invertibility\cite{LinearAlg}.
To invert a system of $m$ equations in $n$ unknowns, with $m > n$, of the form 
(\ref{eq:8times4real}), then
we seek an $n \times m$ matrix ${\mathcal G}$ 
such that  
${\mathcal G} {\mathcal M} = \one $, the unit $n \times n$ matrix, 
and then $A = {\mathcal G}{\mathcal Z}$.  
If such a solution exists, the rank of ${\mathcal M}$ is $n$ ($=4$).

Note that the multiplication of a row of ${\mathcal G}$ with a column of ${\mathcal M}$ 
is actually a Euclidean real inner product.  
If the columns of ${\mathcal M}$ are understood to be spinors over an 8
dimensional real basis, we can accept the same interpretation for the
rows of  ${\mathcal G}$. 
The existence of an 8 dimensional real basis thus supplies 
us with a definition of a real inner product between spinors, 
for which we will use the notation $(\phi \bullet \psi)$. 
It is easy to verify that $(\phi \bullet \psi)$ is actually equal to the real part 
of the standard complex inner product:
\begin{equation}
	(\phi \bullet \chi)=\Re e <\phi , \chi>
	=\frac 12(\phi^\dagger \chi + \psi^\dagger \chi ).
	\label{eq:RealInnerProduct}
\end{equation}

The system of $m$ equations in $n$ unknowns entails\cite{LinearAlg} that 
the right hand side of the equation should satisfy $m-n$ 
($8-4=4$) additional consistency conditions, arising from the fact that ${\mathcal Z}$
must fall in the column space of ${\mathcal M}$.  To find these consistency
conditions, we seek a further $m-n$ linearly independent spinor rows 
$\chi$ that have zero 
real inner product with the columns of ${\mathcal M}$ 
($\chi$ span the left null space of ${\mathcal M}$).
The consistency conditions may then be written $(\chi \bullet Z)=0$.
${\mathcal G}$ is not unique, for any linear combination of the 
rows $\chi$ can be added without changing its effect on ${\mathcal M}$.

It is not necessary to work explicity in 8 real components: 
regardless of which basis we use, the columns of ${\mathcal M}$
and the rows of ${\mathcal G}$ are just spinors in a vector space
over ${\mathbb R}$, and the matrix multiplication is just the
calculation of inner products using (\ref{eq:RealInnerProduct}).
All that we require for the inversion is to find spinors $\phi_\nu$ where 
$(\phi_\nu \bullet \gamma^\mu \psi) \propto \delta^\mu_\nu$.
For the consistency conditions, we require 4 linearly independent spinors 
$\chi$ such that  $(\chi \bullet \gamma^\mu \psi) = 0$.
As will now be shown, the structure 
of the Dirac algebra indeed admits such rows, $\phi_\nu$ and $\chi$.

\subsection*{Inversion}
Let $\phi_\nu = \gamma^{0}\gamma_{\nu} \psi$.  
Then
\[      (\phi_\nu \bullet \gamma^\mu \psi) = \frac 12(
\psi^{\dagger}\gamma_{\nu}^{\dagger}\gamma^{0\; \dagger}\gamma^{\mu}\psi
+ \psi^{\dagger}\gamma^{\mu \; \dagger}\gamma^{0} \gamma_{\nu} \psi).
\]
We use the hermiticity of $\gamma^0$ and $\gamma^0\gamma^\mu$:
\[
\gamma^{\mu\dagger} \gamma^0 = \gamma^{\mu\dagger} \gamma^{0\dagger}
	=  ( \gamma^0 \gamma^\mu)^\dagger = \gamma^0 \gamma^\mu ,
\]
and likewise for $\gamma_\nu$.
Then
\[      (\phi_\nu \bullet \gamma^\mu \psi) = \frac 12(
	\psi^{\dagger}\gamma^{0}(\gamma_{\nu}\gamma^{\mu}
	+ \gamma^{\mu}\gamma_{\nu}) \psi )
={\delta_{\nu}}^{\mu} \overline{\psi}\psi
\label{eq:Deltamunu}
\]
where the Dirac conjugate $\overline{\psi}$ is 
defined in the usual way\footnote{
General properties and nomenclature for the Dirac algebra
in generic dimensions and spacetime signatures (including
four dimensional Minkowski space) are given in \S 3 below.} 
as $\overline{\psi} = \psi^{\dagger}\gamma^{0}$. 

Applying the real inner product with the same rows to the right-hand side 
of the Dirac equation (\ref{eq:DiracEqn}), gives explicitly 
\[ \frac 12(
i\psi^{\dagger}\gamma^0\gamma^\nu \gamma^\mu \partial_\mu \psi -i\partial_\mu 
\psi^{\dagger}\gamma^0 \gamma^\mu \gamma^\nu \psi -
2m \psi^{\dagger }\gamma^0 \gamma ^\nu \psi).
\] 
The last term is identified with the current 
$j^\nu \equiv \overline{\psi} \gamma^\nu \psi$, so 
we can write the solution for the vector potential
\begin{equation}
	A_{\mu}= \frac {1}{2q} \frac{ i(\overline{\psi} \gamma_{\mu} \slash{\partial} \psi
	-  \overline{\psi}\overleftarrow{\slash{\partial}} \gamma_{\mu}\psi) 
	-2 m j_\mu} {\overline{\psi}\psi}.                           
	\label{eq:4DDiracSoln}
\end{equation}

\subsection*{Consistency Conditions}
It is possible to show by the use of (\ref{eq:RealInnerProduct})
two separate sufficient conditions for spinors $\chi$ to have zero real inner product
with $\gamma^\mu \psi$: 
\begin{enumerate}
\item  $\chi=\Gamma \psi$, 
where $\Gamma$ is a matrix in the Dirac algebra such that 
$\Gamma^{\dagger} \gamma^{\mu}$ is 
\textit{anti}hermitean;\\
\emph{Alternatively},
\item $\chi=\Upsilon \psi^{*}$ where $\Upsilon$ is a matrix in the Dirac algebra 
such that $\Upsilon^\dagger \gamma^{\mu}$ is
\textit{antisymmetric}.
\end{enumerate}
As an example of (1), take $\chi=i\gamma^{0} \psi$.  
Both the left hand side as well as the mass term of (\ref{eq:DiracEqn}) 
vanish (as $i\gamma^{0}$ is itself antihermitean), leaving
\[
0 = -\psi^{\dagger}\gamma^0\gamma^\mu \partial_\mu \psi -\partial_\mu
\psi^{\dagger }(\gamma^0\gamma^\mu )^{\dagger }\psi.
\]
This is the normal current conservation equation,
\begin{equation}
 \partial \cdot j \equiv
	\partial_\mu j^\mu = \partial _\mu \psi^{\dagger }\gamma^0\gamma^\mu \psi 
	 =0.
	\label{eq:CurrentCons}
\end{equation}
Also satisfying condition (1), take
$\chi=i\gamma^{0}\gamma_{5}\psi$, using the antihermiticity of 
$i\gamma^{0}\gamma_{5}\gamma^{\mu}$. 
The hermiticity of 
$i\gamma_{5}\gamma^{0}$ now ensures that the mass term survives, and 
manipulations on the right hand side of (\ref{eq:DiracEqn}) lead in a 
similar way to the equation for partial conservation of axial current 
$j_5^\nu \equiv \overline{\psi}\gamma_{5}\gamma^\nu \psi $ 
as the second consistency condition:
\begin{equation}
 \partial \cdot j_5 +2im \overline{\psi}\gamma_{5}\psi =0.
	\label{eq:AxialCurrentCons}
\end{equation}

As an example of the sufficient condition (2), 
take $\chi=\gamma_5 C \psi^{*}$ and $\chi=i \gamma_5 C \psi^{*}$ respectively,  
where $C$ is the charge conjugation matrix. 
We evaluate these inner products with $\gamma ^\mu \psi$ 
using the hermitean conjugate
$(\gamma_5 C \psi^{*})^\dagger = \psi^t (\gamma_5 C)^\dagger
= -\psi^t C \gamma_5 $.
These inner products are the real and imaginary
parts of $ {- \psi^t C \gamma_5 \gamma^\mu \psi} $ respectively, which 
is zero by the antisymmetry of $C\gamma ^5 \gamma ^\mu $.
Applying the same row operation(s) to the right hand side gives
$0=\psi ^tC\gamma ^5(i\gamma ^\mu \partial _\mu \psi -m\psi )$ or
\begin{equation}
	\psi^t C\gamma ^5 \slash{\partial} \psi =0
	\label{eq:4DComplexIdentity}
\end{equation}
(after using the antisymmetry of $C\gamma_5$ itself to eliminate the 
mass term) yielding one complex condition, or two real conditions on the spinor.
This is the result previously reported by Eliezer\cite{Eliezer} 
(who attributed to Dirac the antisymmetry argument using 
$C\gamma_5 = \alpha_x\alpha_z$  in the standard representation).
The consistency conditions (\ref{eq:CurrentCons}),
(\ref{eq:AxialCurrentCons}), and (\ref{eq:4DComplexIdentity}) are 
equivalent to Radford's\cite{Radford96} `reality' conditions.

\subsection*{Alternative inversion}
As mentioned above, the choice of nonsingular matrix inverting
${\mathcal M}$, and consequently the form of the final 
expression for $A$, is not unique.
As an alternative choose $\phi_\nu=i\gamma^{0}\gamma_5\gamma_{\nu} \psi$. 
We then find by a similar working to (\ref{eq:Deltamunu}),
using the anticommuting property of $\gamma_5$ with $\gamma ^\mu$,
that 
\[
(i\gamma^{0}\gamma_5\gamma_{\nu} \psi \bullet \gamma^{\mu}\psi) = 
-\delta_{\nu}^{\mu} \overline{\psi}i\gamma_5\psi. 
\]
Applying the inner product with the same rows to the right hand side 
of the Dirac equation (\ref{eq:DiracEqn}),
in this case the mass term vanishes, yielding an alternative 
solution for the vector potential:
\begin{equation}
	A_\mu =\frac {i}{2q} \frac{
	\overline{\psi }\gamma_5\gamma_\mu \slash{\partial} \psi - 
	\overline{\psi }\gamma_5 \overleftarrow{\slash{\partial}} \gamma_\mu \psi }
	{\overline{\psi }\gamma_5\psi }.
	\label{eq:gamma5inversion}
\end{equation}
That (\ref{eq:4DDiracSoln}) and (\ref{eq:gamma5inversion}) are 
indeed equivalent, and equivalent to \cite{Radford96},
follows from the use of Fierz identities together 
with the auxiliary constraints (see \S 3 below).

\subsection*{$8 \times 5$ real system}
The inversion (\ref{eq:gamma5inversion}) does not contain any mass term.
However, note that the pseudoscalar 
consistency condition (\ref{eq:AxialCurrentCons}) can be written
\begin{equation}
	m=\frac {i}{2} \frac{
	\overline{\psi }\gamma_5 \overleftarrow{\slash{\partial}} \psi 
	+\overline{\psi }\gamma_5 \slash{\partial}  \psi }
	{\overline{\psi }\gamma_5\psi }.
	\label{eq:massinversion}
\end{equation}
The similarity between (\ref{eq:gamma5inversion}) and (\ref{eq:massinversion})
suggests that the original
system could have been considered as 
8 real equations in 5 unknowns, $qA_{0}, \ldots, qA_{3}$, and $m$ (or more generally a Lorentz scalar potential). In this
system, (\ref{eq:gamma5inversion}) and (\ref{eq:massinversion}) provide an inversion, 
while (\ref{eq:CurrentCons}) and the real and imaginary parts of 
(\ref{eq:4DComplexIdentity}) provide the 3 consistency conditions.

\subsection*{2-spinor analysis}
Radford\cite{Radford96} and Booth and 
Radford\cite{BoothRadford} used van der Waerden notation
in order to derive a complex form of the vector potential subject to 
additional reality conditions. Here the 2 spinor version 
is reached via the Weyl representation of the 
Dirac algebra (see for example \cite{ItzyksonZuber}), wherein
\[
\psi_{\alpha} = \left( \begin{array}{c} u_{a} \\ \bar{v}^{\dot{a}} 
\end{array} \right), \nonumber \quad
\psi^{c}_{\alpha}= \left( \begin{array}{c} v_{a} \\ \bar{u}^{\dot{a}} 
\end{array} \right),
	\nonumber  \quad
\overline{\psi}^{\alpha} = 
	    -\left( \begin{array}{c} v^{a} \\ \bar{u}_{\dot{a}} \end{array} \right).
	\label{eq:PsicolumnDef}         
\]
A generic matrix $\Gamma$ in the Dirac algebra has matrix 
elements      
\[
	{\Gamma_{\alpha}}^{\beta} = 
	\left( \begin{array}{cc}{\Gamma_{a}}^{b} & {\Gamma_{a \dot{b}}} \\   
	  {\Gamma^{\dot{a} b}} & {\Gamma^{\dot{a}}}_{ \dot{b}} \end{array} 
	  \right),
	\label{eq:GammaDef}
\]
in particular, 
\[
	{{\gamma^{\mu}}_{\alpha}}^{\beta} = 
	-\left( \begin{array}{cc} 0 & {{\bar{\sigma}}^{\mu}}_{a \dot{b}} \\  
	  {\sigma^{\mu}}^{\dot{a} b} & 0 \end{array} \right) .
	\label{eq:gammamuDef}
\]
where\footnote{The Pauli matrices are
\[
\sigma^{0}=\left( \begin{array}{cc} 1 & 0 \\ 0 & 1 \end{array} \right), 
\nonumber       \quad
\sigma^{1}=\left( \begin{array}{cc} 0 & 1 \\ 1 & 0 \end{array} \right), 
\nonumber       \quad
\sigma^{2}=\left( \begin{array}{cc} 0 & -i \\ i & 0 \end{array} \right), 
\nonumber       \quad 
\mbox{and} \nonumber       \quad
\sigma^{3}=\left( \begin{array}{cc} 1 & 0 \\ 0 & -1 \end{array} \right).      
\]
}
\begin{equation}
	({\sigma}^{\mu})_{0 \le \mu \le 3}=
	(\sigma^{0}, \mbox {\boldmath $\sigma$}), 
	\quad ({\bar{\sigma}}^{\mu})_{0 \le \mu \le 3}=
	(\sigma^{0}, -\mbox {\boldmath $\sigma$}).          
	\label{eq:SigmaDefns}
\end{equation}
The definitions (\ref{eq:SigmaDefns}) are consistent with
\[
	C_{\alpha \beta} = -\left( \begin{array}{cc} \varepsilon_{ab} & 0 \\ 
	 0 & \varepsilon^{\dot{a} \dot{b}} \end{array} \right),
	\quad
	C^{\alpha \beta} = -\left( \begin{array}{cc} \varepsilon^{ab} & 0 \\  
	 0 & \varepsilon_{\dot{a} \dot{b}} \end{array} \right)
	\label{Cdef}
\]
together with $\varepsilon = i\sigma^{2}$, that is, component-wise,
\[
\varepsilon_{ab}=\left( \begin{array}{cc} 0 & 1 \\ -1 & 0 \end{array} 
\right) = \varepsilon_{\dot{a}\dot{b}},
\quad \varepsilon^{a b} = - \varepsilon_{a b}, \quad 
\varepsilon^{\dot{a}\dot{b}} = - \varepsilon_{\dot{a}\dot{b}}.
	\label{eq:EpsilonSet}
\]

Starting then from
\[
	q \gamma^{\mu}A_{\mu}\psi = (i \gamma^{\mu}\partial_{\mu} - m) \psi
	\label{Diracequn}
\]
and transcribing to 2-spinor form, the Dirac equation 
reads directly
\[
	q \bar{A}_{a \dot{b}} \bar{v}^{\dot{b}}   =  i \bar{\partial}_{a 
	\dot{b}} \bar{v}^{\dot{b}} + m u_{a}, \label{eq:Upper'} \quad
	 q A^{\dot{a}b}u_{b} =  i \partial^{\dot{a}b}u_{b} + m 
	 \bar{v}^{\dot{a}} \label{eq:Lower'}
\]
where\footnote{Hermiticity, and raising and lowering of indices are 
entailed in the relations
${{\bar{\sigma}}^{\mu}}_{a \dot{a}} = \varepsilon_{ab}\varepsilon_{\dot{a}\dot{b}}
{\sigma^{\mu}}^{\dot{b} b}$, 
$({\sigma^{\mu}}^{\dot{a} b})^{*}={\sigma^{\mu}}^{\dot{b} a}$.
}
\[
	{\bar{A}}_{a \dot{b}} \equiv {\bar{\sigma}^{\mu}}_{a \dot{b}}A_{\mu}, 
	\quad
	A^{\dot{a} b} = {{\sigma}_{\mu}}^{\dot{a} b} A^{\mu}.
	\label{eq:4vector2spinor}
\]
Finally taking complex conjugates,
\begin{eqnarray}
U: \quad q A^{\dot{d} c}\bar{v}_{\dot{d}} & = 
&  -i \partial^{\dot{d} c}\bar{v}_{\dot{d}} + m u^{c}, \nonumber  \\
\bar{V}: \quad q A^{\dot{a} b} u_{b}
& = & i \partial^{\dot{a} b} u_{b} + m \bar{v}^{\dot{a}},  \nonumber \\
\bar{U}: \quad q A^{\dot{c} d} v_{d}
& = & i \partial^{\dot{c} d} v_{d} +m 
\bar{u}^{\dot{c}}, \nonumber \\
V: \quad q A^{\dot{b} a} \bar{u}_{\dot{b}}
& = & -i \partial^{\dot{b} a} 
\bar{u}_{\dot{b}} + m v^{a}. \nonumber
\end{eqnarray}
Thus by taking combinations of the form 
$\alpha( \bar{V}^{\dot{a}} v^{b} - \bar{U}^{\dot{a}} u^{b})$,
	$\beta ( U^{b} \bar{u}^{\dot{a}} -  {V}^{b} \bar{v}^{\dot{a}})$ and using 
$u_{d}v^{b}-v_{d}u^{b} = {\delta_{d}}^{b}(u_{c}v^{c})$
the vector potential can be isolated, with general `solution'
\begin{equation}
A^{\dot{c}d}= -\frac{i}{q}
\left[  \frac { \alpha( v^{d} \partial^{\dot{c}e} u_{e} + u^{d} \partial^{\dot{c}e} v_{e}) - 
\beta (\bar{u}^{\dot{c}} \partial^{\dot{e}d} \bar{v}_{\dot{e}}- \bar{v}^{\dot{c}}\partial^{\dot{e}d}\bar{u}_{\dot{e}})
	- 2 i m ( \alpha u^{d}\bar{u}^{\dot{c}} + \beta  v^{d}\bar{v}^{\dot{c}})     }
	{(\alpha u^{a}v_{a} + \beta \bar{u}^{\dot{a}} \bar{v}_{\dot{a}})} 
	\right] .
	\label{eq:1parcNumberSoln2spinorForm}
\end{equation}
with arbitrary parameters $\alpha, \beta$.
As emphasised by Radford\cite{Radford96}, all these forms are 
equivalent, subject 
to the hermiticity conditions satisfied by the potential itself. The 
latter can be imposed via the two-spinor projections of $A^{\dot{a} 
b}$, namely
\begin{eqnarray}
	(A^{\dot{a} b}\bar{u}_{\dot{a}}u_{b})^{*}&=& (A^{\dot{a} 
	b}\bar{u}_{\dot{a}}u_{b}),  \nonumber \\
	(A^{\dot{a} b}\bar{v}_{\dot{a}}v_{b})^{*}&=& (A^{\dot{a} 
	b}\bar{v}_{\dot{a}}v_{b}),  \nonumber \\
	(A^{\dot{a} b}\bar{u}_{\dot{a}}v_{b})^{*}&=& (A^{\dot{a} 
	b}\bar{v}_{\dot{a}}u_{b}).  \label{eq:AHermiticity}
\end{eqnarray}
Substitution of these conditions into a suitable form of (\ref 
{eq:1parcNumberSoln2spinorForm}), for example with $\alpha=1,\;
\beta=0$, leads to
\begin{eqnarray}
	\partial_{\mu}(v^{a}{{\bar{\sigma}}^{\mu}}_{a \dot{b}}\bar{v}^{\dot{b}}+ 
	\bar{u}_{\dot{a}}{\sigma^{\mu}}^{\dot{a} b} u_{b})& = & 0,
	\nonumber  \\
	\partial_{\mu}(v^{a}{{\bar{\sigma}}^{\mu}}_{a \dot{b}}\bar{v}^{\dot{b}}+ 
	\bar{u}^{\dot{a}}{\sigma^{\mu}}^{\dot{a} b} u_{b})& = & 
	2im(v^{a}u_{a} - \bar{u}_{\dot{a}}\bar{v}^{\dot{a}}),
	\nonumber  \\
	u^{a}{{\bar{\sigma}}^{\mu}}_{a \dot{b}}\partial_{\mu}\bar{v}^{\dot{b}} 
	& = & \bar{v}_{\dot{a}}{\sigma_{\mu}}^{\dot{a} b}\partial^{\mu}u_{b}.
	\label{eq:2SpinorConstraints}
\end{eqnarray}
Referring (\ref{eq:1parcNumberSoln2spinorForm}) to a fixed tetrad basis, 
$\bar{u}_{\dot{a}}u_{b}$, $\bar{v}_{\dot{a}}v_{b}$, 
$\bar{u}_{\dot{a}}v_{b}$ and $\bar{v}_{\dot{a}}u_{b}$ via appropriate 
contractions and using (\ref {eq:2SpinorConstraints}) then shows 
directly that all forms are equivalent to the manifestly hermitean 
version with $\alpha=\beta =1$. As expected from the general analysis 
in the previous subsection, (\ref{eq:1parcNumberSoln2spinorForm}) with $\alpha=\beta =1$
agrees with the previous result (\ref{eq:4DDiracSoln}) expressed in the Weyl 
basis, and (\ref{eq:2SpinorConstraints}) are equivalent to 
current and partial axial current conservation\footnote{
Take the sum and difference of (\ref{eq:2SpinorConstraints}a) and 
(\ref{eq:2SpinorConstraints}b)} and the additional complex 
pseudoscalar identity (\ref{eq:4DComplexIdentity}) satisfied by the Dirac wavefunction.   

\section{Higher dimensional extensions}
In the previous section it was shown that the 4 dimensional Dirac 
equation can be regarded as a singular set of real linear equations for 
the vector potential $A_{\mu}$. Gaussian elimination in this $8 
\times 4$ real system then leads to a solution for $A_{\mu}$
and also implies a set of four additional 
linearly independent constraints, linear in $\partial_{\mu}\psi$, 
which can be identified in this case with bilinear identities proposed 
by Radford\cite{Radford96} and Booth and Radford\cite{BoothRadford} 
using van der Waerden 2-spinor notation. Either approach ultimately 
derives from the structure of the Dirac algebra and the symmetry 
properties of the $\gamma$ matrices.
The same analysis is extended here in Dirac notation to higher dimensional cases
and different metric signatures (in flat spacetime). Also, an important 
distinction to make is that between $c$-number 
and $a$-number (or Grassmann-valued) Dirac spinors. For the latter, a 
Gaussian elimination argument requires a formal treatment of linear 
algebra over Grassmann-extended ground fields\cite{deWitt};  
for present purposes it suffices to assume that the count of 
solutions and constraints goes through.

The explicit construction of the Dirac algebra in higher dimensions 
has been reviewed by Tanii\cite{Tanii}; see also \cite{Townsend, 
Morette,Coquereau}. The Dirac matrices satisfy
\begin{equation}
	\{ \gamma^{\mbox{}}_{\mu}, \gamma^{\mbox{}}_{\nu} \} = 
	2 \eta^{\mbox{}}_{\mu \nu},
	\label{eq:DiracAlg}
\end{equation}
where the spacetime metric signature is taken as $(t,s)$ for even 
dimensions $t+s=d$, indices being labelled $\mu,\nu = 0, 0', 0'', \ldots; 
1, 2, \ldots, s$,
with chirality determined by the equivalent of $\gamma_{5}$, $\widehat{\gamma} \equiv
\gamma^{\mbox{}}_{0} \gamma^{\mbox{}}_{0'} \ldots 
\gamma^{\mbox{}}_{1} \gamma^{\mbox{}}_{2}\ldots\gamma^{\mbox{}}_{d}$.

There are three involutive automorphisms associated 
respectively with complex conjugation, transposition and hermitean 
conjugation under which the $\gamma$ matrices acting on complex spinors 
$\psi_{\alpha}$ undergo similarity transformations\footnote{The notation 
${\gamma^{(a)}_{\mu}}$, ${\gamma^{(b)}_{\mu}}$, ${\gamma^{(c)}_\mu}$ is 
used
for the expressions on the right hand sides of 
(\ref{eq:Conjugations}) (without the sign factors), 
extended to arbitrary elements of the Dirac algebra $\Gamma$ (see 
below). Note that in the previous section the index conventions for 
the $A$ and $C$ 
matrices differ from that used  here in the general case.},
\begin{eqnarray}
	{{\gamma^{\mbox{}}_{\mu}}_{\alpha}}^{\beta} & = & \delta_{A}\; A_{\alpha 
	\alpha'}\;
	{{\gamma^{\dagger}_{\mu}}^{\alpha'}}_{\beta'}\; (A^{-1})^{\beta' \beta}  \\
	{{\gamma^{\mbox{}}_{\mu}}_{\alpha}}^{\beta} & = & \delta_{B}\; 
	{B_{\alpha}}^{\alpha'} \;
	{{\gamma^{*}_{\mu}}_{\alpha'}}^{\beta'}\; {(B^{-1})_{\beta'}}^{\beta}  \\
	{{\gamma^{\mbox{}}_{\mu}}_{\alpha}}^{\beta} & = & \delta_{C}\; C_{\alpha \alpha'}\;
	{{\gamma^{T}_{\mu}}^{\alpha'}}_{\beta'}\; (C^{-1})^{\beta' \beta} 
	\label{eq:Conjugations}
\end{eqnarray}
where $\delta_{A}$, $\delta_{B}$, $\delta_{C}$ are sign factors 
depending on the spacetime:
\begin{eqnarray*}
	\delta_{A} & = & \delta_{B}\delta_{C} \; ;  \\
	\delta_{B} & = & \pm 1 \; ;  \\
	\delta_{C} & = & \delta_{B}(-1)^{t+1}.
	\label{eq:SigmaSigns}
\end{eqnarray*}
$A$, $B$ and $C$ are related through the definitions of Dirac and 
charge conjugation for spinors,
\begin{eqnarray*}
	{\psi_{c}}_{\alpha} & = & {B_{\alpha}}^{\beta} {\psi^{*}}_{\beta}  \\
	{\overline{\psi}}^{\alpha}  & = & {\psi^{*}}_{\beta} (A^{-1})^{\beta  \alpha}   \\
	{\psi^{c}}_{\alpha} & = & C_{\alpha \beta}{\overline{\psi}}^{\beta}
\end{eqnarray*}
so that 
$$
{B_{\alpha}}^{\beta} = C_{\alpha \beta'}(A^{-1})^{\beta \beta'}.
$$

The fundamental identity $B^{*}B = 
\epsilon_{B} \one$ determines the existence of Majorana spinors for metrics 
in which $\epsilon_{B}=+1$, and $\delta_{B}=-1$. Dimensions for which 
this is possible can be read from table \ref{tbl:EpsilonB}
which gives the values of $\epsilon_{B} = \sqrt{2} \cos[\frac 14 \pi 
(s-t - \delta_{B})]$. Finally, the implementation of parity in the 
Dirac algebra is determined by the symmetry of $C$, $C^{T} = 
\epsilon_{C} C$, $\epsilon_{C} = (\delta_{B})^{t}(-1)^{\frac 12 
t(t-1)}\epsilon_{B} $.

\begin{table}[tbp]
	\centering
	\caption{The values of $\epsilon_{B}$ as a function of $\delta_{B}= 
	\pm 1$ and $s-t \; (\mbox{mod}\; 8)$ (from \cite{Tanii}; $\times$ 
	indicates that no representation with the specified signs 
	exists).}\vspace*{.2cm}
	\begin{tabular}{|c||c|c|c|c|c|c|c|c|}
		\hline 
		$s-t$ & 0 & 1 & 2 & 3 & 4 & 5 & 6 & 7  \\
		\hline \hline
		$\epsilon_{+}$ & +1 & $\times$ & $-1$ & $-1$ & $-1$ & $\times$ & +1 & +1  \\
		\hline
		$\epsilon_{-}$ & +1 & +1 & +1 & $\times$ & $-1$ & $-1$ & $-1$ & $\times$  \\
		\hline
	\end{tabular}
	
	\label{tbl:EpsilonB}
\end{table}
   
Consider now the Dirac equation for $\psi$ and  
conjugate forms:
\begin{eqnarray}
   q\slash{A} \psi & = & (i\slash{\partial} -m)\psi \nonumber \\
	q\slash{A} \psi^{c} & = & (-i\slash{\partial} -\delta_{B}m)\psi^{c} \nonumber   \\
	q\overline{\psi}\slash{A} & = & \overline{\psi}(-i\overleftarrow{\slash{\partial}} -\delta_{A}m) 
	\nonumber \\
	q\overline{\psi^{c}}\slash{A} & = & \overline{\psi^{c}}(i\overleftarrow{\slash{\partial}} -\delta_{C}m) 
	\label{eq:DiracEqForms}
\end{eqnarray}
where the sign factors follow from the explicit definitions in 
(\ref{eq:Conjugations}). Write (\ref {eq:DiracEqForms})
as $\Psi$, $\overline{\Psi}$, $\Psi^{c}$
and $\overline{\Psi^{c}}$ respectively. Then, as in the four dimensional case, the 
solution for the potential follows by taking combinations
$\overline{\psi}\gamma^{\mbox{}}_{\mu}\Psi + 
\overline{\Psi}\gamma^{\mbox{}}_{\mu}\psi$
which force the isolation of $A_{\mu}$ through an anticommutator of 
$\gamma$ matrices:
\begin{equation}
	A_{\mu}= \frac {1}{2q} \frac{ i(\overline{\psi} \gamma_{\mu} \slash{\partial} \psi
	-  \overline{\psi}\overleftarrow{\slash{\partial}} \gamma_{\mu}\psi) 
	-2s_{A}m \overline{\psi} 
	\gamma_{\mu}\psi} {\overline{\psi}\psi}.                           
	\label{eq:PotSoln}
\end{equation}
where $s_{A}= \frac 12 (1+ \delta_{A}) \equiv \frac 12 (1-(-1)^{t})$ 
(compare (\ref{eq:4DDiracSoln})).

The identification of additional identities satisfied by the Dirac wavefunction given 
(\ref{eq:DiracEqForms}), (\ref{eq:PotSoln}) amounts to determining the 
structure of a certain ideal in the free algebra of rational 
expressions in $\psi$, $\psi^{*}$ and partial derivatives 
$\partial^{\mu} \psi$,
$\partial^{\mu} \psi^{*}$, for appropriately smooth wavefunctions. 
Such expressions 
do not necessarily form linear representations of the spacetime Lorentz 
symmetry group $SO(t,s)$, and the
distinguished role played by the constraints 
following from Gaussian elimination is not clear. 
Here, the quadratic case is analysed, by analogy with the four-dimensional 
case, and in relation to the counting suggested by dimensional considerations.
 \textit{Polynomials} in $\psi$, $\overline{\psi}$ 
and derivatives \textit{do} decompose with respect to the Lorentz 
algebra. Moreover, there is a natural bi-grading by degree:
in the quadratic case, this is simply by fermion number $\mathbb F$.
Thus the cases $\overline{\psi} \cdot \psi$, with ${\mathbb F} = 0$, 
and 
$\psi \cdot \psi$, or equivalently
$\overline{\psi^{c}} \cdot \psi$, with $\mathbb F =2$,   
can be considered (as the latter will necessarily be complex, 
constraints in $\overline{\psi} \cdot \overline{\psi}$ do not 
require separate treatment). A case by case 
analysis follows.
\begin{description}
	\item[${\mathbb F} = \pm 2$]  :\\
	As is well known, products of two spinors admit a 
	decomposition into antisymmetric tensor representations of the Lorentz 
	group\cite{KingWybourne}. Noting that  
	\begin{equation}
		\overline{\psi^{c}} \Gamma \psi \equiv \epsilon_{B} (C^{-1} 
		\Gamma)^{\alpha \beta} \psi_{\alpha} \psi_{\beta}, 
		\label{eq:BilinearPsiSymm}
	\end{equation}
	it is clear that an explicit decomposition is provided by the
	linear basis for the Dirac algebra comprising the antisymmetric $p$-fold products of $\gamma$ 
		matrices\cite{Tanii, Delbourgo} $\{ \gamma_{\mu}, \gamma_{\mu \nu}, 
		\ldots, \gamma_{\mu_{1}\mu_{2}\ldots \mu_{p}}, \ldots \}$, $p=1, \ldots, 
		d$. Furthermore, for appropriate $p$, depending on the spacetime 
		signature, dimension and statistics of the spinor wavefunctions, (\ref{eq:BilinearPsiSymm}) is 
		identically zero, leading after contraction with $A^{\mbox{}}_{\mu}$ 
		to a sequence of differential identities in the Dirac wavefunction, 
		after implementation of (\ref{eq:DiracEqForms}, \ref{eq:PotSoln}). If the 
		statistics of the Dirac wavefunction is specified as
		$\psi_{\alpha}\psi_{\beta} = f \psi_{\beta}\psi_{\alpha}$, then 
		using
	\[
		\overline{\psi^{c}} \Gamma \psi = \epsilon_{C}f \overline{\psi^{c}} 
		\Gamma^{(c)} \psi, 
		\label{eq:BilinearShuffle}
	\]
	and
	\begin{equation}
		(\gamma_{\mu \mu_{1}\mu_{2}\ldots \mu_{p}})^{(c)} =  
		\delta_{C}^{p+1}\gamma_{\mu_{p} \ldots \mu_{2} \mu_{1} \mu}
	=  \delta_{C}^{p+1}(-1)^{\frac 12(p+1)p} \gamma_{\mu \mu_{1}\mu_{2}\ldots \mu_{p}},
		\label{eq:BilinearShuffleSign} 
	\end{equation}
	we require
	\begin{equation}
		\epsilon_{C}f\delta_{C}^{p+1}(-1)^{\frac 12(p+1)p} = -1.
		\label{eq:BilinearPsiVanish}
	\end{equation}
	Note also\cite{Delbourgo}
	\begin{eqnarray}
	\gamma_{\mu} \gamma_{ \mu_{1}\mu_{2}\ldots \mu_{p}} & = &
	(-1)^{p} \gamma_{\mu_{1}\mu_{2}\ldots \mu_{p}}\gamma_{\mu} + 
	2p\eta_{\mu[\mu_{1}} \gamma_{\mu_{2}\ldots \mu_{p}]} \nonumber \\ 
	 & = & -(-1)^{p} \gamma_{\mu_{1}\mu_{2}\ldots \mu_{p}}\gamma_{\mu} + 
	{\textstyle \frac 2{(p+1)}} \gamma_{\mu\mu_{1}\mu_{2}\ldots \mu_{p}}. 
		\label{eq:SimplifyHighRank}
	\end{eqnarray}
    \begin{theo}
		For spacetime dimension $d= 4k$, a single Lorentz scalar or 
		pseudoscalar bilinear complex differential constraint of the form
		\begin{equation}
			\overline{\psi^{c}} \slash{A} \psi =0 \quad \mbox{or} \quad 
			\overline{\psi^{c}} \widehat{\gamma} \slash{A} \psi =0
			\label{eq:BilinConstraints}
		\end{equation}
		exists, whenever $\epsilon_{C}\delta_{C}f = \mp 1$, respectively.
		For spacetimes of dimension $d= 4k+2$, both scalar and pseudoscalar complex 
		constraints (\ref{eq:BilinConstraints})
		hold if $\epsilon_{C}\delta_{C}f = -1$; otherwise neither holds. 
		In general a sequence of Fierz type identities
		\begin{equation}
			 \delta_{p}\overline{\psi^{c}} \gamma_{\mu_{1}\mu_{2}\ldots \mu_{p}} \slash{A} \psi 
			  = p A_{[\mu_{1}}\overline{\psi^{c}} \gamma_{\mu_{2}\ldots \mu_{p}]}\psi ,
			\label{eq:FierzType}
		\end{equation}
		holds, where $\delta_{p}=
		\frac 12 (1-\epsilon_{C}\delta_{C}^{p+1} f(-1)^{\frac 12 p(p+1)})$. 
		Note that 
		(\ref{eq:BilinConstraints}), (\ref{eq:FierzType}) are regarded as 
		differential conditions on $\psi$ via the 
		substitutions (\ref{eq:DiracEqForms}), 
		(\ref{eq:PotSoln}).
	\end{theo}
	\begin{proof}\\
		Clearly (\ref{eq:SimplifyHighRank}), together with the interchange 
		sign factors implies (\ref{eq:FierzType}) after contraction with 
		$A^{\mbox{}}_{\mu}$. After substituting (\ref{eq:PotSoln}) and 
		rationalising, the identity is therefore quartic, and  
		generically of Fierz type. 
	    The only bilinear cases occur when $p=0$ (in which 
	    case the condition $\epsilon_{C}\delta_{C}f = -1$ refers to the 
	    symmetry of $\overline{\psi^{c}} \slash{A} \psi$), or for $p=d$
		(for which $\gamma_\mu$ and $\widehat{\gamma}$ anticommute, and the 
		charge conjugation of $\widehat{\gamma}$ determines whether the 
		pseudoscalar identity holds). 
	\end{proof}
	   \item[${\mathbb F} = 0$]  :
	\begin{theo}
	The real type Fierz type identities
		\begin{equation}
		\overline{\psi} \slash{A} \gamma_{\mu_{1}\mu_{2}\ldots \mu_{p}}  \psi = (-1)^{p}
		\overline{\psi} \gamma_{\mu_{1}\mu_{2}\ldots \mu_{p}} \slash{A} \psi
			  + 2 p A_{[\mu_{1}}\overline{\psi} \gamma_{\mu_{2}\ldots \mu_{p}]}\psi 
			\label{eq:FierzTypeR}
		\end{equation}
		hold, where the appropriate parts of (\ref{eq:DiracEqForms})
		are to be used on the left and right-hand sides, respectively.  
		Current conservation and partial 
		conservation of axial current hold for all d:
		\begin{eqnarray}
			\partial_{\mbox{}}^{\mu} \overline{\psi} \gamma^{\mbox{}}_{\mu} \psi 
			&=& 0, \nonumber \\
			\partial_{\mbox{}}^{\mu} \overline{\psi} \widehat{\gamma} \gamma^{\mbox{}}_{\mu} \psi 
			+ 2ims_A (\overline{\psi} \widehat{\gamma}  \psi) &=& 0 .
			\label{eq:PCCnAC}
		\end{eqnarray}
		\end{theo}
	    \begin{proof}\\
	    (\ref{eq:FierzTypeR}) follows directly by contraction of 
	    (\ref{eq:SimplifyHighRank}a) with $A_\mu$.
	    (\ref{eq:PCCnAC}) uses (\ref{eq:DiracEqForms}) together with 
	    anticommutativity of $\gamma_\mu$ and $\widehat{\gamma}$ 
	    (See (\ref{eq:SimplifyHighRank}b)).
	    \end{proof}
	\end{description}
Explicit enumeration of these results is complicated because 
of the multitude of subcases involved. In table \ref{tbl:DiverseDimns}
the counting and nature of the identities is illustrated in diverse dimensions, and 
for given fermion statistics and metric signatures for 
the Minkowski, conformal and Euclidean spacetimes. The 
explicit expressions for the sign factors in the representations of 
the Dirac algebra available imply the expressions given for 
$\delta_{C}\epsilon_{C}$, namely $\varepsilon_{B}$, 
$\delta_{B}\varepsilon_{B}$ and $(-1)^{\frac 12 
d}\delta_{B}\varepsilon_{B}$ for $t=1$, $t=2$ and $t=d$ respectively.
The entries within each metric signature class then indicate for 
which type of fermion wavefunction statistics ($c$-number, $f=+1$, 
or $a$-number, $f=-1$ respectively) the indicated scalar 
$S$ or pseudoscalar $P$ complex identity exists. Where 
$\varepsilon_{B}=+1$ exists, 
bracketed entries indicate the choice $\delta_{B}=-1$ or 
$\delta_{B}=+1$,
consistent with the availability either of Majorana $(M)$ or 
pseudoMajorana $(pM)$ 
spinors respectively, in that dimension and spacetime signature. 
\begin{table}[pb]
	\centering
	\caption{{\small Enumeration of the type 
		of complex bilinear identity (scalar, $S$ or pseudoscalar, $P$) admitted by the Dirac wavefunction of 
		the indicated statistics, for various dimensions $d$ and metrics of 
		Minkowski $(1,d-1)$, conformal $(2,d-2)$ and Euclidean $(d,0)$ 
		signature
		(see text for details). }}\vspace*{.2cm}
	\begin{tabular}{|c|c||c|c|c|c||c|c|c|c||c|c|c|c|}
		\hline
		\multicolumn{2}{|c||}{\mbox{}} & \multicolumn{4}{c||}{$(1,d-1)$} & 
		\multicolumn{4}{c||}{$(2,d-2)$} & 
		\multicolumn{4}{c|}{$(d,0)$}  \\
		\cline{3-14}
		\multicolumn{2}{|c||}{\mbox{}} & $\;\;\varepsilon_{B}\;\;$ & $S$ & $P$ &  & 
		$\;\delta_{B}\varepsilon_{B}\;$ & 
		 $S$ & $P$ &  & $\pm \delta_{B}\varepsilon_{B}$ & $S$ & 
		 $P$ &   \\
		\hline
		& 2 & +1 & $(a)$ & $(a)$ & $(M)$ & -- & -- & -- & -- & $-1$ & $(c)$ & $(c)$ & $(M)$  \\
		\cline{2-14}
		& 4 & $\mp 1$ & $(a)$ & $(c)$ & $(M)$ & $\pm 1$ & $(c)$ & $(a)$ & 
		$(M)$ & $\pm 1$ & $\begin{array}{c}a \\[-.15 cm] c \end{array}$ & 
		$\begin{array}{c}c \\[-.15 cm] a \end{array}$ & --  \\
		\cline{2-14}
		$d$ & 6 & $-1$ & $c$ & $c$ & -- & $-1$& $(c)$ & $(c)$ & $(M)$ & +1 & $(a)$ & 
		$(a)$ & $(pM)$  \\
		\cline{2-14}
		& 8 & $\pm 1$ & $(a)$ & $(c)$ & $(pM)$ & $\mp 1$ & $\begin{array}{c}c 
		\\[-.15 cm] a \end{array}$
		& $\begin{array}{c}a \\[-.15 cm] c \end{array}$ & -- & $\mp 1$ & $(a)$ & 
		$(c)$ & $(M)$  \\
		\cline{2-14}
		&10 & +1 & $(a)$ & $(a)$ & $(M)$ & +1 & $(a)$ & $(a)$ & $(pM)$ & 
		$-1$ & $(c)$ & $(c)$ & $(M)$  \\
		\hline
	\end{tabular}
	\label{tbl:DiverseDimns}
\end{table}

In table \ref{tbl:DiverseDimns}, the 
entry for $c$-number wavefunctions in four dimensional 
Minkowski space corresponding to a single complex bilinear 
pseudoscalar identity is the original Dirac equation case. For $a$-number fermions, there exists the 
corresponding scalar equivalent (reflecting the properties of 
$\gamma_{\mu}$ and $\gamma_{\mu}\gamma_{5}$ in the Dirac algebra).
Taking into account the two real constraints (current and partial axial current 
conservation), there is in either case a total of four real 
conditions, in accord with the count needed to accompany the four 
dimensional vector potential in the $8 \times 4$ linear system (see \S 
2).

The counting of constraints in $d$ dimensions must be examined 
carefully in relation to Gaussian elimination in the corresponding 
$2^{\frac 12 d +1}\times d$ real linear system. For example in 
two-dimensional Minkowski space there are \textit{no} complex (${\mathbb 
F} = \pm 2$) bilinear 
constraints for $c$-number wavefunctions, and thus only two real (${\mathbb 
F} = 0$) bilinear constraints, whereas for $a$-number wavefunctions there are  
two real plus \textit{two} complex identities. The apparent under- or 
over- determination of the system (at the quadratic level) must be reconciled with the 
remaining Fierz-type identities (of quartic type). Two 
dimensions is a special case because of the abelian nature 
of the Lorentz group, but in other cases, the Lorentz decomposition 
of the remaining higher order constraints also bears on the counting.
For example, in $d=6$ ($16 \times 6$ real system) there are again either no or two complex 
bilinear constraints (see table \ref{tbl:DiverseDimns}). In the 
latter case, a further $16-6-2-4=4$ independent real conditions are 
needed. These may represent 4 
additional scalar conditions at higher order, or a real 
6-vector which satisfies two additional conditions equivalent to two 
of the scalar and pseudoscalar conditions. Similar considerations 
apply to the $d=8$ and $d=10$ cases.

\section{Conclusions}
In this work the linearity of the Dirac equation has been exploited as 
a vehicle to obtain an algebraic solution for the vector potential, 
and previous discussions in the literature 
(\cite{Radford96} and \cite{Eliezer}) have been reconciled. In 
addition to its role in the problem of obtaining (classical) solutions of the full 
nonlinear Maxwell-Dirac equations, the algebraic method has potential 
application to the nonabelian case, and to Duffin-Kemmer algebras rather 
than Clifford algebras. Modifications to electrodynamics such as the 
Born-Infeld theory, and indeed the nonrelativistic limit of the Dirac 
equation itself, may also be amenable to further study by the method.

\subsubsection*{Acknowledgements}
The authors are grateful to Prof Angas Hurst for discussions, and for 
providing copies of his own notes, and to Prof Christie Eliezer for informing 
one of us (HSB) of his paper and original 
correspondence with Dirac.  We also thank Profs Tony Bracken and Bob Delbourgo 
for interest in the work and discussions of various aspects. Finally PDJ thanks Wim Tholen for 
inspiration.

\end{document}